\documentclass[lettersize,journal]{IEEEtran}
\usepackage{amsmath,amsfonts}
\usepackage{algorithmic}
\usepackage{algorithm}
\usepackage{array}
\usepackage[caption=false,font=normalsize,labelfont=sf,textfont=sf]{subfig}
\usepackage{textcomp}
\usepackage{stfloats}
\usepackage{url}
\usepackage{verbatim}
\usepackage{graphicx}
\usepackage{cite}
\usepackage{siunitx}

\begin{document}

\title{Deep Diffusion Deterministic Policy Gradient based Performance Optimization for Wi-Fi Networks}

\author{
\IEEEauthorblockN{Tie~Liu, Xuming~Fang, Rong~He}

\IEEEauthorblockA{Key Laboratory of Information Coding and Transmission, Southwest Jiaotong University, Chengdu, China\\
tliu@my.swjtu.edu.cn, xmfang@swjtu.edu.cn, rhe@swjtu.edu.cn}
        % <-this % stops a space
\thanks{The work was supported in part by NSFC under Grant 62071393. (Corresponding author: Xuming Fang)}
}



\maketitle

\begin{abstract}
Generative Diffusion Models (GDMs), have made significant strides in modeling complex data distributions across diverse domains. Meanwhile, Deep Reinforcement Learning (DRL) has demonstrated substantial improvements in optimizing Wi-Fi network performance. Wi-Fi optimization problems are highly challenging to model mathematically, and DRL methods can bypass complex mathematical modeling, while GDMs excel in handling complex data modeling. Therefore, combining DRL with GDMs can mutually enhance their capabilities. The current MAC layer access mechanism in Wi-Fi networks is the Distributed Coordination Function (DCF), which dramatically decreases in performance with a high number of terminals. In this study, we propose the Deep Diffusion Deterministic Policy Gradient (D3PG) algorithm, which integrates diffusion models with the Deep Deterministic Policy Gradient (DDPG) framework to optimize Wi-Fi network performance. To the best of our knowledge, this is the first work to apply such an integration in Wi-Fi performance optimization. We propose an access mechanism that jointly adjusts the contention window and the aggregation frame length based on the D3PG algorithm. Through simulations, we have demonstrated that this mechanism significantly outperforms existing Wi-Fi standards in dense Wi-Fi scenarios, maintaining performance even as the number of users increases sharply.
\end{abstract}


\section{Introduction}
\IEEEPARstart{A}{s} Wi-Fi technology evolves, the rapid development of advanced network applications such as 8K video streaming, augmented/virtual reality (AR/VR), and remote surgery poses unprecedented challenges. These demands have significantly surpassed the capabilities of existing standards such as 802.11ax, driving the development of the next generation of Wi-Fi technologies, particularly IEEE 802.11be (Wi-Fi 7). These new technologies aim to deliver extremely high throughput (EHT) and reduced latency to meet the complex requirements of future applications.

Meanwhile, traditional mathematical modeling approaches struggle with the stochastic and variable nature of wireless channels and the complex interdependencies among parameters. In this paper, we explore the optimization of the MAC layer access mechanism by jointly adjusting the contention window and aggregation frame length, which are interdependent parameters. Longer aggregation frames can significantly reduce transmission overhead; however, they may also lead to higher retransmission overhead and higher packet error rate, especially in dense deployed scenarios. This issue can be mitigated either by increasing the contention window (CW) or by reducing the aggregation frame length. Recent advances in machine learning, especially reinforcement learning algorithms, offer effective solutions by bypassing complex mathematical modeling steps through their adaptive learning capabilities and data processing ability. These algorithms are increasingly pivotal in optimizing Wi-Fi networks to meet changing demands.

In recent years, Generative Artificial Intelligence (GAI) has made remarkable strides across various domains\cite{croitoru2023diffusion}. GAI technologies excel at creating innovative content by analyzing large datasets that include text, images, and music. Particularly in the field of data augmentation\cite{trabucco2023effective}, generative models synthesize additional data to bolster machine learning models, providing critical support in scenarios where data is scarce or privacy concerns are paramount. For network optimization tasks, such as the joint optimization of interrelated parameters, GAI proves invaluable in handling highly complex data distributions and capturing intricate relationships. Nevertheless, the challenge lies in determining the optimal combination of parameter adjustments for various environments.

Du \textit{et al}. \cite{du2023beyond} introduced how generative diffusion models can be applied to network optimization problem, providing an easy example of power optimization, which shows promising performance in sum rate compared to DRL methods. In another work of Du \textit{et al}. \cite{du2024diffusion}, they address the increasing demand for efficient content generation capabilities in the next generation of Internet with the emergence of the Metaverse. They introduce the Artificial Intelligence Generative Optimal Decision (AGOD) algorithm, which combines DRL and diffusion models, enabling the application of generative diffusion theory to network performance optimization.

The Distributed Coordination Function (DCF), a core Medium Access Control (MAC) mechanism in current Wi-Fi standards, operates on the basis of Carrier Sense Multiple Access with Collision Avoidance (CSMA/CA) and Binary Exponential Backoff (BEB). DCF follows a listen-before-talk principle: Before transmitting, the device checks whether the channel is idle. If busy, it waits for a random backoff period within $[0,CW_{max}]$, where the contention window doubles after each collision. Once the channel becomes free, the device transmits its data.

Although effective in low-density scenarios, BEB performance deteriorates as the number of stations (STAs) increases. Bianchi \cite{bianchi2000performance} demonstrated that dense STAs lead to higher collision probabilities, packet losses, and reduced throughput. This limitation poses significant challenges for future high-density Basic Service Set (BSS) environments, where excessive contention severely impacts network performance.

In Wi-Fi transmission, frame aggregation is an effective technique used to enhance network performance. The length of aggregated frames can impact latency, channel utilization, throughput, etc. Therefore, it is necessary to study how to optimize the length of aggregated frames under different channel conditions and varying Quality of Service (QoS) requirements. Camps-Mur \textit{et al.} \cite{camps2012leveraging} discuss how to utilize the frame aggregation mechanism in the 802.11 standard to improve QoS in Wi-Fi networks. Coronado \textit{et al.} \cite{coronado2020adaptive} proposed an adaptive machine learning-based approach for optimizing frame size selection by considering channel conditions, demonstrating an average goodput improvement of 18.36\% over standard aggregation mechanisms. Zhou \textit{et al.} \cite{10333585} propose a Joint Frame Length and Rate Adaptation (JFRA) scheme based on Double Deep Q-learning (DDQN), which outperforms the Minstrel HT and Thompson Sampling algorithm by up to 21.3\% and 68.9\% in various cases.

This paper proposes a joint optimization approach of adjusting the CW and aggregation frame length to enhance throughput in Wi-Fi networks with densely distributed terminals. The relationship between the optimal CW and aggregation frame length for each user is extremely complex, so simultaneous adjustments for each terminal necessitate a model or algorithm capable of effectively modeling these complex relationships. This represents a significant challenge in optimizing network performance while accommodating the high density of terminals. We apply a method named D3PG which combines generative diffusion models (GDMs) and deep deterministic policy gradient (DDPG). D3PG melds the strengths of the DDPG with the sophisticated modeling capabilities of GDMs.

The rest of the paper is organized as follows. Section \ref{SEC:2} presents the system model. Section \ref{SEC:3} introduces how to apply diffusion models into DDPG. Section \ref{SEC:4} gives the performance of D3PG. The simulation and analysis are based on NS3 platform\cite{ns3website}. In the end
, we summarize this paper in Section \ref{SEC:5}.

\section{System Model}\label{SEC:2}
\begin{figure}[!ht]
  \centering
    \includegraphics[width=0.45\textwidth]{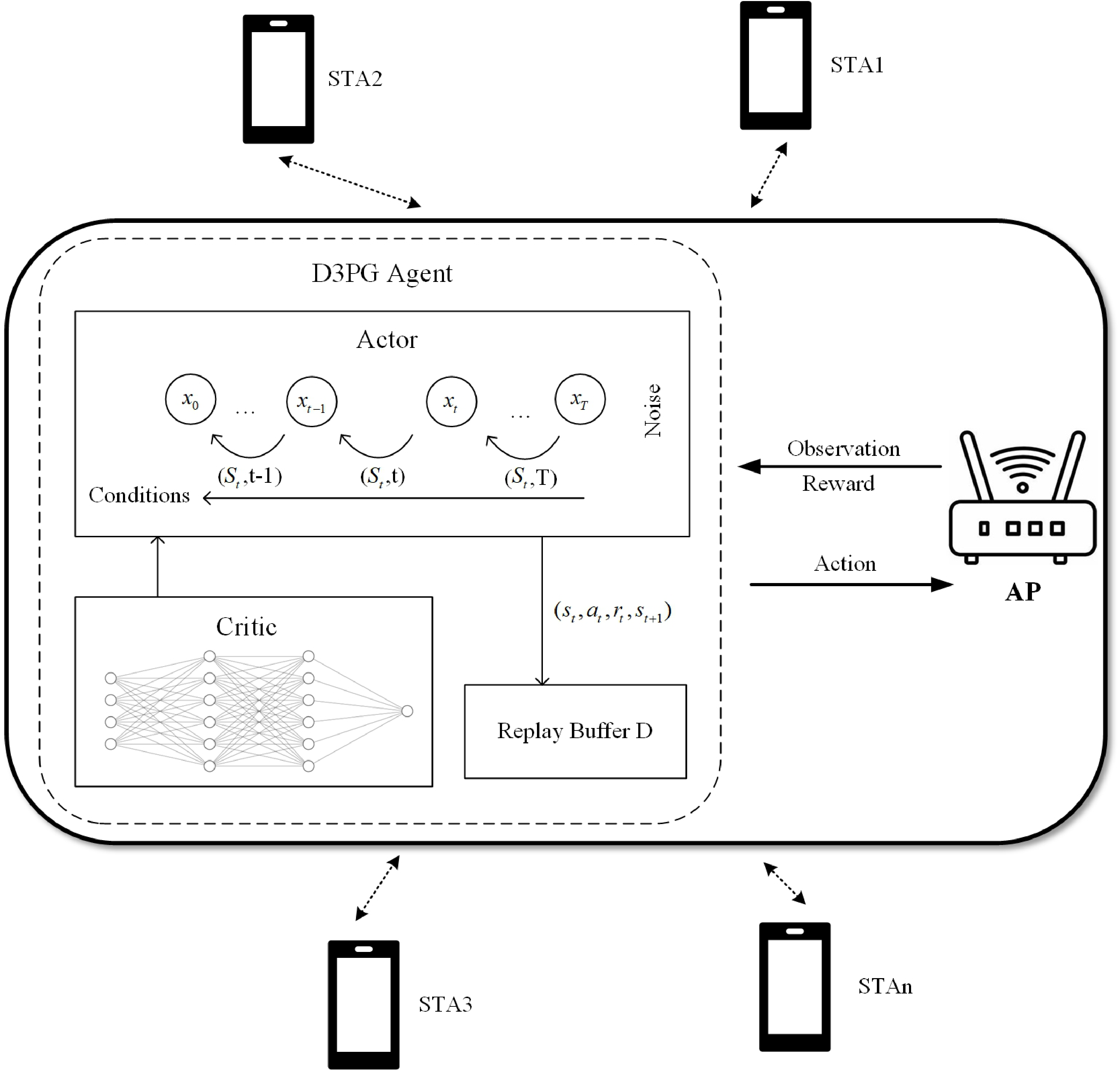}
    \caption{System Model}
    \label{Figure:p1}
\end{figure}

Our system model is illustrated in Fig.~\ref{Figure:p1}, which includes one Access Point (AP) and $N_s$ Stations (STAs), which are randomly distributed within a circle of radius $r$ around the AP. The D3PG agent is located at the AP, which gathers all the necessary data for training the model, either directly or through reports from the STAs. Our objective is to train the model to optimize the total throughput in dense Wi-Fi scenarios.

The main difference between DDPG and D3PG is the modeling of policy network, while the optimization process can also be regarded as a Markov Decision Process (MDP)\cite{qiu2020nndl} that requires the model's agent to collect state, action, and reward information. During training, the agent at the AP collects the state information $s_t$, and $s_t=\{ITP_t, PLR_{t,i}|i\in N_s\}$ is consisted of channel Idle Time Proportion (ITP) and Packet Loss Rate (PLR). ITP is the percentage that the AP detects the channel is idle in a certain period, which is calculated as below:
\begin{equation}
    ITP_t=\frac{t_{Idle}}{t_{period}}
\end{equation}
PLR indicates the reliability of data transmission, which is calculated by the number of transmitted frames and the number of received ACK frames $n_{ACK, i}$ in a certain period:
\begin{equation}
    PLR_{t,i}=1-\frac{n_{TX, i}}{n_{ACK, i}}
\end{equation}
where $n_{TX, i}$ and $n_{ACK, i}$ represent separately the number of transmitted frames and received ACK frames for the $i$th STA. Then the D3PG agent, starting from the acquired state $s_t$, processes through a conditioned diffusion model along with randomly generated noise matching the action space size to get the optimal action. After multiple denoising steps, the model outputs an action $a_t=\{CW_{t, i}, L_{t, i}|i\in N_s\}$, which consists of contention window $CW_{t, i}$ and aggregation frame length $L_{t, i}$ for each STA. The model directly outputs continuous numbers between 0 and 1, and we discretely map $CW_{t, i}$ to values between $CW_{min}$ and $CW_{max}$, and $L_{t, i}$ to values between $L_{min}$ and $L_{max}$.
The AP then issues the action to STAs, executing it for a certain period, results in a reward $r_t$ and the next state $s_{t+1}$. This completes one trajectory $\tau_t=(s_t, a_t, r_t, s_{t+1})$ which is stored in an experience pool $D$ for future model optimization.

The reward function is designed to optimize these adjustments based on the improvement in throughput.
\begin{equation}
    r_t=2\times (sigmoid(\frac{Thr_t}{\lambda})-0.5)
\end{equation}
where $Thr_t$ represents the throughput of the whole system at time $t$, $\lambda$ is a hyperparameter that controls the sensitivity of the reward to throughput. Considering practicality, and $\lambda$ can be set to a desired throughput or the maximum Shannon capacity. This allows us to normalize any positive real number representing throughput to a range between 0 and 1, thus accurately reflecting reward information and aiding in model convergence.

In this approach, D3PG enables the AP to learn optimal policies for complex network environments, thereby enhancing overall network performance in scenarios that are increasingly prevalent in modern Wi-Fi applications.

\section{Design of Optimization scheme}\label{SEC:3}
The D3PG algorithm is based on the deep deterministic policy gradients algorithm, with the primary distinction being in replacing the decision network by a conditioned diffusion model\cite{du2024diffusion}. The generative diffusion model can be divided into two parts: the forward diffusion process and the reverse diffusion process\cite{ho2020denoising}.

\subsection{Forward process}

In computer vision, $x_t$ represents an image, whereas in our network optimization context, $x_t$ denotes a solution which is CW or aggregation frame length setting for each STA at time $t$. The following equation shows how noise diffuses in a solution.

\begin{equation}
    x_{t+1} = \sqrt{\beta}\times \epsilon + \sqrt{1-\beta}\times x_t
    \label{eq:fp}
\end{equation}
where $\epsilon$ is Gaussian noise with the same dimensions as $x_t$, and $\beta$ is a noise diffusion coefficient, which is a scalar between $0$ and $1$. $x_{t+1}$ is the result of $x_t$ plus the noise. The value of $\beta$ starts near zero at the beginning and gradually increases close to $1$. This process represents the diffusion of noise into the original data, with the gradual increase in $\beta$ indicating that the speed of diffusion is accelerating.

Following a series of derivations, we get the following equation:
\begin{equation}
    x_t = \sqrt{1-\bar\alpha}\times \epsilon + \sqrt{\bar\alpha}\times x_0
    \label{eq:xt=x0}
\end{equation}
where $x_0$ is an optimal solution, $\bar\alpha=\alpha_t\alpha_{t-1}\dots\alpha_1$, and $\alpha_t=1-\beta_t$, where $\alpha_t$ is defined to simplify the equations. So we get the relationship between $x_0$ and $x_t$. 
This completes the derivation of the forward diffusion process for the generative diffusion model.

\subsection{Reverse process}
The reverse diffusion process involves deriving the relationship from a known $x_t$ back to $x_0$. Similar to the forward process, we first derive the relationship for one step of reverse diffusion, which is $P(x_{t-1} | x_t)$. According to Bayes' theorem, we have:
\begin{equation}
    P(x_{t-1}|x_t) = \frac{P(x_{t}|x_{t-1})P(x_{t-1})}{P(x)}
    \label{eq:bayes}
\end{equation}
Using (\ref{eq:fp}) and (\ref{eq:xt=x0}) derived from the forward diffusion process and substituting and rearranging terms in (\ref{eq:bayes}), we find that $P(x_{t-1} | x_t)$ follows the following normal distribution:

\begin{equation}
    P(x_{t-1}|x_t)\sim 
    N(\mu, \sigma^2)
\end{equation}

\begin{equation}
    \mu=\frac{\sqrt{\alpha_t}(1-{\bar\alpha_{t-1}})}{1-\bar{\alpha_t}}x_t
    +\frac{\sqrt{\bar\alpha_{t-1}}(1-{\alpha_{t}})}{1-{\alpha_t}}\frac{x_t-\sqrt{1-\bar\alpha}\times\epsilon}{\sqrt{\bar\alpha}} \label{eq:mu}
\end{equation}

\begin{equation}
    \sigma=\sqrt{\frac{\sqrt{1-\alpha_t}\sqrt{1-\bar\alpha_{t-1}}}{\sqrt{1-\bar{\alpha_t}}}}
\end{equation}

In (\ref{eq:mu}), $\epsilon$ represents the noise added directly to an optimal solution $x_0$ at any given moment $t$. Knowing the noise $\epsilon$ added from $x_0$ to $x_t$ allows us to determine the probability distribution of the previous moment $x_{t-1}$. This is the essence of the reverse diffusion process.

\begin{algorithm}[htb]
\caption{D3PG: Deep Diffusion Deterministic Policy Gradient}\label{alg:D3PG}
\begin{algorithmic}[1]
\STATE Initialize replay buffer $D$, critic $Q(\theta^Q)$, and conditioned diffusion model actor $\mu(\theta^\mu)$;
\STATE Initialize target network with the same weights.
\STATE Define $\Delta t$ as interaction period.
\STATE Define the maximum training time step $T_{max}$.
\STATE Define the maximum denoising steps $\tau_{max}$.

\FOR{$t = 1$ to $T_{max}$}
    \STATE Get $s_t$ from environment.
    \STATE Sample Gaussian noise $a_N$.
    \FOR{$\tau = \tau_{max}$ to $1$}
        \STATE Infer $a_t$ from conditioned diffusion model $\mu(\theta^\mu)$ based on $s_t$, $a_N$, and $\tau$.
    \ENDFOR
    \IF{training}
        \STATE $a_t \leftarrow a_t + noise_t$.
        \STATE Execute $a_t$ and observe $r_t$ and $s_{t+1}$.
        \STATE Store transition $\tau_t=(s_t, a_t, r_t, s_{t+1})$ in $D$.
        \STATE Sample a random minibatch of $N$ transitions $batch_N = \{(s_i, a_i, r_i, s_{i+1})|i\in N\}$ from $D$.
        \STATE Update critic $Q(\theta^Q)$ and $\mu(\theta^\mu)$ based on $batch_N$.
        \STATE Soft update the target network.
    \ENDIF
    \STATE Execute $a_t$.
\ENDFOR
\end{algorithmic}
\end{algorithm}
A neural network can be trained, with inputs including the noised solution $x_t$, to predict the noise $\epsilon$ that has been added to the optimal solution $x_0$. This predicted noise allows us to obtain the probability distribution of the solution at the previous moment. Sampling randomly from this distribution produces a solution of the previous moment, which can then be fed back into the network to predict its noise, and this process is repeated until the desired solution $x_0$ is obtained.

By training a neural network to continuously perform conditioned reverse diffusion, where environmental information is input together with the noised solution $x_t$, final $x_0$ obtained is the result of optimizing using the generative diffusion model. In the implementation of D3PG, only the reverse process of the diffusion model is applied, as the decision network in DDPG is replaced by the diffusion model.

\section{Performance Evaluation}\label{SEC:4}
To evaluate the effectiveness of the D3PG algorithm, we implemented it using PyTorch and designed the Wi-Fi scenario on the NS3 platform, an open-source, discrete-event network simulation tool. To integrate machine learning algorithms into the Wi-Fi scenario and dynamically adjust multiple network parameters, we utilized the NS3-AI module\cite{10.1145/3389400.3389404}. NS3-AI serves as an interface between NS3 and PyTorch, facilitating the application of intelligence techniques within the simulation environment of NS3. 

The architecture of the simulated Wi-Fi network is depicted in Fig. \ref{Figure:p1}, designed based on the 802.11ax standards. The parameters utilized in the simulation are detailed in Table \ref{table:ns3-param}. D3PG hyperparameter setting is in table \ref{table:hyperparameters}.

\begin{table}[h]
  \caption{NS3 Simulation Parameters}
  \label{table:ns3-param}
  \centering
  \begin{tabular}{c|c}
 %  \begin{tabular}{|p{2.5cm}|p{4cm}|}
    \textbf{Parameter} &  \textbf{Value} \\
    \hline
    Number of STAs & $\{8, 16, 24, 32, 40, 48, 56, 64\}$\\
    % \hline
    Frequency & 5GHz\\
    % \hline
    Max A-MSDU size  &6160 bytes\\
    % \hline
    Max A-MPDU size& 1586176 bytes\\
    % \hline
    $CW_{min}, CW_{max}$ & $15,1023$ \\
    % \hline
    $L_{min}, L_{max}$& $1,256$ \\
    % \hline
    Distance between STAs and AP & In the circle with radius $r=7.5$(m)\\
    % \hline
    Simulation time & 100s\\
    % \hline 
    Loss model &LogDistancePropagationLossModel
    \\
    % \hline 
    Traffic arrival model  &Constant rate model, rate=1Gbps\\
    % \hline
    Channel bandwidth  & 80MHz\\
    % \hline
    Number of spatial streams  &2\\
    % \hline
    payload size  &1448 bytes\\
    % \hline
    traffic source  &TCP traffic\\
    % \hline
    Error rate model  &TableBasedErrorRateModel\\
    % \hline
    Mobility & Stationary \\ 
    % \hline
    \end{tabular}
  \end{table}

\begin{table}[h]
    \caption{D3PG Hyperparameters}
    \label{table:hyperparameters}
    \centering
    \begin{tabular}{c|c}
        % \begin{tabular}{|p{2.5cm}|p{4cm}|}
        \textbf{Parameter} &  \textbf{Value} \\
        \hline
        % \hline
        Interaction period with NS3 $\Delta t$ & $50ms$\\
        Hidden layer dimension & 256 \\ %\hline
        % Interaction period & 50ms \\ %\hline
        Reward normalization coefficient $\lambda$& 450 \\ %\hline
        Actor learning rate & \num{2e-3} \\ %\hline
        Critic learning rate & \num{2e-2} \\ %\hline
        Soft update factor $\tau$ & \num{5e-2} \\ %\hline
        Reward discount factor $\gamma$ & \num{1e-1} \\ %\hline
        Batch size & 12 \\ %\hline
        Buffer size & 256 \\ %\hline
        $\beta$ schedule & Variance Preserving\\ %\hline
        Denoise steps & 5\\ %\hline
        
    \end{tabular}
\end{table}

In our study, we first compared the throughput performance of our algorithm in a scenario with 64 STAs. We selected three different algorithms as baseline algorithms for performance comparison. They are:

\begin{itemize}
    \item Baseline 1: Utilizes the existing 802.11 Wi-Fi standard's MAC layer contention access mechanism, which is based on BEB and CSMA/CA, with fixed frame aggregation length.
    \item Baseline 2: Proximal Policy Optimization (PPO) algorithm proposed by Schulman \textit{et al.} \cite{schulman2017proximal}, which is an on-policy reinforcement learning algorithm.
    \item Baseline 3: DDPG algorithm that does not integrate generative diffusion models, with the same hyperparameter settings as our D3PG algorithm.
\end{itemize}

Specifically, Baseline 2 and Baseline 3 use the same states, actions, and reward functions as our proposed method.

% The hardware utilized in our simulation platform primarily includes a CPU (Intel Core i5-13600KF) and a GPU (NVIDIA RTX 4060). Simulations were conducted in a Linux environment, specifically Ubuntu version 22.04.4 LTS.

It should be noted that due to the randomness of Wi-Fi networks, we conducted five repeated simulations for each algorithm and took the average to minimize the impact of randomness on the comparative results. 

% \begin{table}[htbp]
% \centering
% \caption{Simulation time comparison}
% \label{tab:simulation_times}
% \begin{tabular}{c|c}

% \textbf{Simulation} & \textbf{Time(min)} \\ \hline
% Baseline 1       & 15                      \\ 
% Baseline 2       & 26.4                      \\ 
% Baseline 3       & 24.2                      \\ 
% D3PG             & 23.4                      \\

% \end{tabular}
% \end{table}

% Table \ref{tab:simulation_times} presents the time required to simulate 100 seconds of real-world activity for each scenario. Baseline 1 represents the simulation of the defined Wi-Fi scenario using only the NS3 platform, while the other three simulations integrate NS3 with NS3-AI, applying different algorithms. As shown, Baseline 1, without data exchange with NS3-AI, demonstrates lower simulation time. In contrast, the time differences among the NS3-AI-based scenarios are minimal, indicating that the computational overhead of AI algorithms is negligible compared to the overall simulation load of the Wi-Fi environment.

% In practical Wi-Fi scenarios, the computational complexity of the algorithms employed is significantly lower than that of diffusion models commonly used in image processing. This is primarily due to the substantially reduced dimensionality of states and actions in Wi-Fi networks compared to images. Moreover, the computation time required by these algorithms is negligible relative to the actual operational times of Wi-Fi networks in real-world environments.

\begin{figure}[!ht]
    \centering
    \vspace{-0.3cm}
    \includegraphics[width=0.3\textwidth]{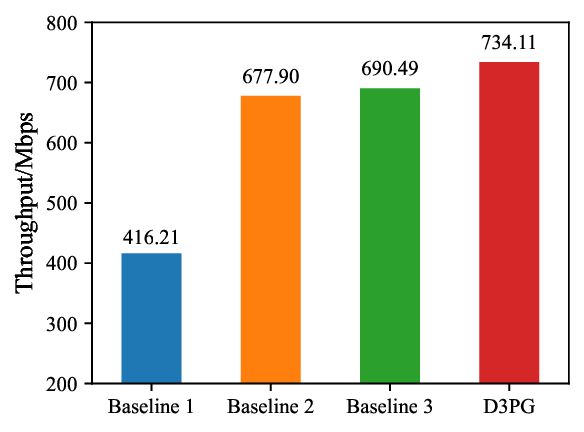}
    \caption{Total Average Throughput Comparison of Different Algorithm}
    \label{Figure:p2}
\end{figure}

As can be seen from Fig. \ref{Figure:p2}, the D3PG algorithm demonstrated a throughput enhancement of 74.6\% over baseline 1, it also showed an additional increase of 13.5\% in throughput compared to baseline 2, and an additional increase of 10.5\% over baseline 3. 

\begin{figure}[!ht]
    \centering
    \vspace{-0.3cm}
    \includegraphics[width=0.3\textwidth]{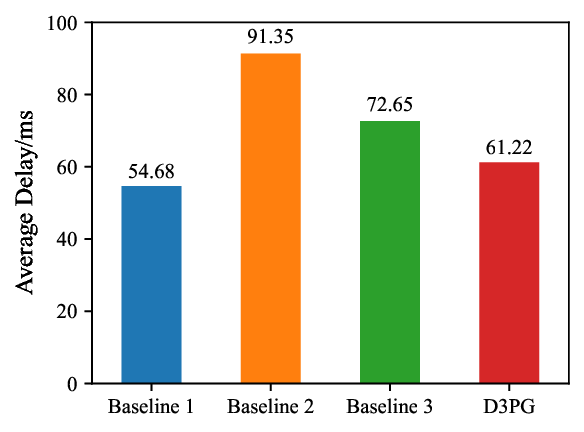}
    \caption{Average Delay Comparison of Different Algorithm}
    \label{Figure:p3}
\end{figure}

In terms of latency performance, as shown in Fig. \ref{Figure:p3}, the D3PG algorithm exhibits an average latency increase of only 6.54 ms compared to Baseline 1, while outperforming both Baseline 2 and Baseline 3. Specifically, compared to baseline 3, the primary reason for the better performance is the aid of generative diffusion model, the D3PG algorithm can more accurately map the relationships between environmental information and action combinations, resulting in more precise parameter control in complex scenarios.

\begin{figure}[!ht]
    \centering
    \vspace{-0.3cm}
    \includegraphics[width=0.3\textwidth]{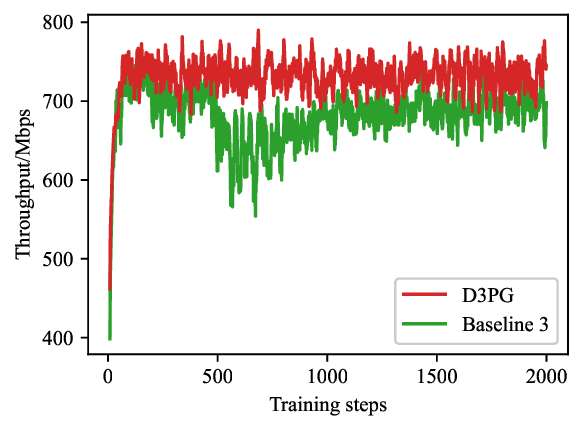}
    \caption{Training Process D3PG vs. Baseline 3}
    \label{Figure:p4}
\end{figure}

Fig. \ref{Figure:p4} shows the changes in total throughput during the training process of the D3PG and baseline 3 algorithms. The convergence times of different algorithms differ,
%; therefore, each algorithm was simulated for a uniform duration of 100 seconds, which was deemed sufficient for all algorithms to achieve convergence. 
the D3PG algorithm training process is observed to be more stable, while the baseline 3 algorithm exhibits greater fluctuations, and D3PG can achieve higher throughput, which means better performance. This is because D3PG has a stronger exploratory ability, which allows the model to reach optimum performance more quickly and reduces training time. This feature gives D3PG another advantage because, in rapidly changing Wi-Fi environments, if the model requires retraining due to environmental changes, the D3PG algorithm can achieve faster convergence, thus minimizing the impact on user experience.

\begin{figure}[!ht]
    \centering
    \vspace{-0.3cm}
    \includegraphics[width=0.3\textwidth]{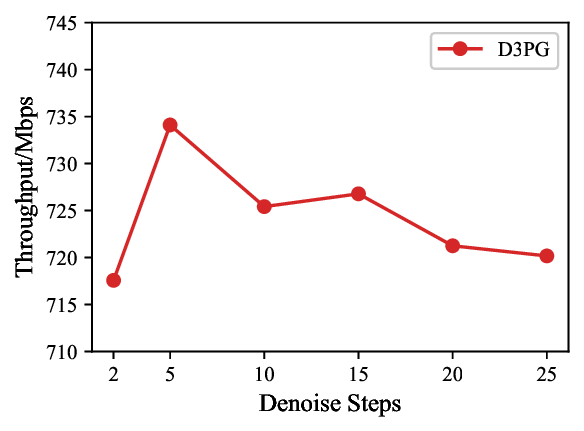}
    \caption{Comparison of Different Denoise Steps}
    \label{Figure:p5}
\end{figure}

As shown in Fig. \ref{Figure:p5}, the number of diffusion steps impacts the performance of the D3PG algorithm. When the number of diffusion steps is too small, the model’s inference is insufficient, leading to suboptimal performance. Conversely, an excessive number of diffusion steps increases computational resource demands and may result in model overfitting, ultimately degrading performance.

\begin{figure}[!ht]
    \centering
    \vspace{-0.3cm}
    \includegraphics[width=0.3\textwidth]{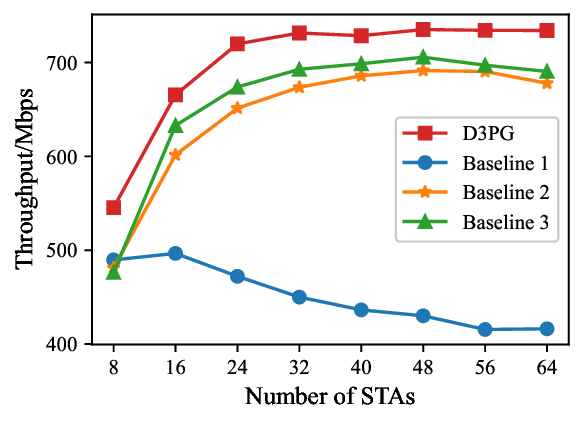}
    \caption{Throughput Across Networks of Different Number of STAs}
    \label{Figure:p6}
\end{figure}

Finally, we evaluated the throughput performance of our algorithm against baselines in networks with varying numbers of STAs. As shown in Fig. \ref{Figure:p6}, all STAs are uniformly distributed within a circle centered on the AP with a radius of 7.5 meters. Our algorithm consistently improves performance across all network sizes. For Baseline 1, throughput initially increases with the number of users due to higher traffic but eventually declines as intensified competition within the BSS leads to collisions, highlighting the limitations of the BEB mechanism in dense scenarios. In contrast, the D3PG algorithm achieves stable throughput as the network saturates. While throughput increases with user count initially, it stabilizes without significant decline, even under heavy competition, thanks to the optimizations introduced by our algorithm.

\section{Conclusion}\label{SEC:5}
In this paper, we introduce a MAC layer access mechanism based on the D3PG algorithm, which enhances network throughput by jointly adjusting the contention window and frame aggregation for each STA. This mechanism addresses the sharp decline in throughput performance caused by increased competition within BSS in dense Wi-Fi scenarios. The throughput improvements with D3PG optimization are significant, showing a 76.4\% increase over the baseline 1 and surpassing baseline 2 and baseline 3 by 13.5\% and 10.5\%, respectively. Besides, supported by generative diffusion models, which excel at modeling complex data distributions, the D3PG algorithm offers more stable training, faster convergence, and greater flexibility and adaptability. We initially applied GAI to optimize the MAC layer performance of Wi-Fi networks, and the results indicate that this approach effectively enhances performance.

% \IEEEpubidadjcol
 

%

\bibliography{IEEEabrv,myreferences}

\bibliographystyle{IEEEtran}

\end{document}